\documentclass[multi]{cambridge7A}
\usepackage[UKenglish]{babel}
\usepackage[longnamesfirst,sectionbib]{natbib}
\usepackage{chapterbib}
\usepackage{amsmath,amssymb,amsthm,amsfonts}
\usepackage{enumerate,url}
\usepackage{graphicx}
\setcitestyle{authoryear,round,semicolon}
\allowdisplaybreaks[1]

\theoremstyle{plain}
\newtheorem*{nthm}{Theorem}

\def\argmin{\mathop{\rm argmin}}
\def\a{\alpha}
\def\bias{{\rm bias}}
\def\da{\downarrow}
\def\decon{_{{\rm decon}}}
\def\ep{\epsilon}
\def\etal{{\em et al.}~}
\def\etalc{{\em et al.},~}
\def\ha{{\hat a}}
\def\hatc{{\hat c}}
\def\hf{{\hat f}}
\def\hg{{\hat g}}
\def\half{^{1/2}}
\def\j{^{(j)}}
\def\ka{\kappa}
\def\kl{_{k\ell}}
\def\mi{\,|\,}
\def\mo{^{-1}}
\def\mt{^{-2}}
\def\mhf{^{-1/2}}
\newcommand{\oonh}{\frac1{nh}}
\newcommand{\ootp}{\frac1{2\pi}}
\def\ra{\to}
\def\rai{\to\infty}
\def\sumi{\sum_i\,}
\def\sumion{\sum_{i=1}^n\,}
\def\sumjon{\sum_{j=1}^n\,}
\def\tf{{\tilde f}}
\newcommand{\thf}{{\textstyle \frac12}}
\def\var{{\rm var}}

\providecommand*\Index[1]{#1\index{#1}}
\providecommand*\undex[1]{} 
\providecommand*\Undex[1]{#1} 

\begin{document}
\alphafootnotes
\author[Aurore Delaigle and Peter Hall]{Aurore
  Delaigle\footnotemark[1]
  and Peter Hall\footnotemark[2]}
\chapter[Empirical deconvolution]{Kernel methods and minimum contrast
  estimators for empirical deconvolution}
\footnotetext[1]{Department of Mathematics and Statistics, The University of
  Melbourne, Parkville, VIC 3010, Australia; A.Delaigle@ms.unimelb.edu.au}
\footnotetext[2]{Department of Mathematics and Statistics, The University of
  Melbourne, Parkville, VIC 3010, Australia; halpstat@ms.unimelb.edu.au}
\arabicfootnotes
\contributor{Aurore Delaigle
  \affiliation{University of Melbourne and University of Bristol}}
\contributor{Peter G. Hall
  \affiliation{University of Melbourne and University of California at Davis}}
\renewcommand\thesection{\arabic{section}}
\numberwithin{equation}{section}
\renewcommand\theequation{\thesection.\arabic{equation}}
\numberwithin{figure}{section}
\renewcommand\thefigure{\thesection.\arabic{figure}}

\begin{abstract}We survey classical kernel methods for providing nonparametric
  solutions to problems involving measurement error.  In particular we outline
  kernel-based methodology in this setting, and discuss its basic properties.
  Then we point to close connections that exist between kernel methods and
  much newer approaches based on minimum contrast techniques.  The connections
  are through use of the sinc kernel for kernel-based inference.  This
  `infinite order' kernel is not often used explicitly for kernel-based
  deconvolution, although it has received attention in more conventional
  problems where measurement error is not an issue.  We show that in a
  comparison between kernel methods for density deconvolution, and their
  counterparts based on minimum contrast, the two approaches give identical
  results on a grid which becomes increasingly fine as the bandwidth
  decreases.  In consequence, the main numerical differences between these two
  techniques are arguably the result of different approaches to choosing
  smoothing parameters.
\end{abstract}

\subparagraph{Keywords}bandwidth, inverse problems, kernel estimators, local
  linear methods, local polynomial methods, minimum contrast methods,
  nonparametric curve estimation, nonparametric density estimation,
  nonparametric regression, penalised contrast methods, rate of convergence,
  sinc kernel, statistical smoothing

\subparagraph{AMS subject classification (MSC2010)}62G08, 62G05

\section{Introduction}
\subsection{Summary}
Our aim in this paper is to give a brief survey of kernel methods\index{kernel methods|(} for solving problems involving \Index{measurement error}, for example problems involving density \Index{deconvolution} or regression with errors in variables\index{errors in variables|(}, and to relate these `classical' methods (they are now about twenty years old) to new approaches based on \Index{minimum contrast methods}.  Section~1.1 motivates the treatment of problems involving errors in variables, and section~1.2 describes conventional kernel methods for problems where the extent of measurement error is so small as to be ignorable.  Section~2.1 shows how those standard techniques can be modified to take account of measurement errors, and section~2.2 outlines theoretical properties of the resulting estimators.

In section~3 we show how kernel methods for dealing with measurement error are
related to new techniques based on minimum contrast ideas.  For this purpose,
in section~3.1 we specialise the work in section~2 to the case of the sinc
kernel\index{kernel methods!sinc kernel}.  That kernel choice is not widely used for density deconvolution,
although it has previously been studied in that context by Stefanski and
Carroll (1990)\index{Stefanski, L.|(}\index{Carroll, R. J.|(}, Diggle and Hall
(1993)\index{Diggle, P. J.}, Barry and Diggle (1995)\index{Barry, J.}, Butucea
(2004)\index{Butucea, C.}, Meister (2004)\index{Meister, A.} and Butucea and
Tsybakov (2007a,b)\index{Tsybakov, A. B.}.  Section~3.2 outlines some of the
properties that are known of sinc kernel estimators, and section~3 points to
the very close connection between that approach and minimum contrast, or
penalised contrast, methods\index{penalised contrast methods}.

\subsection{Errors in variables}
Measurement errors\index{measurement error|(} arise commonly in practice, although only in a minority of statistical analyses is a special effort made to accommodate them.  Often they are minor, and ignoring them makes little difference, but in some problems they are important and significant, and we neglect them at our peril.

Areas of application of deconvolution\index{deconvolution|(}, and regression with measurement
error, include the analysis of seismological\index{seismology} data
(e.g.~Kragh and Laws, 2006)\index{Kragh, E.}\index{Laws, R.}, \Index{financial
analysis} (e.g.~Bonhomme and Robin, 2008)\index{Bonhomme, S.}\index{Robin, J.-M.},
disease epidemiology\index{epidemiology|(} (e.g.~Brookmeyer and Gail, 1994,
Chapter~8)\index{Brookmeyer, R.}\index{Gail, M. H.}, and nutrition\undex{nutrition|(}.

The latter topic is of particular interest today, for example in connection
with errors-in-variables problems for
\index{data!nutrition data|(}data gathered in
\index{food|(}food frequency
questionnaires\index{questionnaire} (FFQs), or dietary questionnaires for
epidemiological studies (DQESs).  Formally, an FFQ is `A method of dietary
assessment in which subjects are asked to recall how frequently certain foods
were consumed during a specified period of time,' according to the Nutrition
  Glossary of the European Food Information Council.  An FFQ seeks detailed
information about the nature and quantity of food eaten by the person filling
in the form, and often includes a query such as, ``How many of the above
servings are from fast food outlets (McDonalds, Taco Bell, etc.)?'' (Stanford
University, 1994)\index{Stanford University}.  This may seem a simple question to answer, but nutritionists interested in our consumption of fat generally find that the quantity of fast food that people admit to eating is biased downwards from its true value.  The significant concerns in Western society about fat intake, and about where we purchase our oleaginous food, apparently influences our truthfulness when we are asked probing questions about our eating habits.

Examples of the use of statistical deconvolution in this area include the work
of Stefanski and Carroll (1990)\index{Stefanski, L.|)}\index{Carroll, R. J.|)} and
Delaigle and Gijbels (2004b)\index{Gijbels, A.}, who address nonparametric
density deconvolution from measurement-error data, obtained from FFQs during
the second National Health and Nutrition Examination Survey (1976--1980);
Carroll \etal(1997)\index{Freedman, L. S.}\index{Pee, D.}, who discuss design
and analysis aspects of linear measurement-error models when data come from
FFQs; Carroll \etal(2006)\index{Midthune, D.}\index{Kipnis, V.}, who use
measurement-error models, and deconvolution methods, to develop marginal mixed
measurement-error models for each nutrient in a nutrition study, again when
FFQs are used to supply the data; and Staudenmayer \etal(2008)\index{Ruppert,
D.}\index{Buonaccorsi, J. P.}, who employ a dataset from nutritional epidemiology\index{epidemiology|)} to illustrate the use of techniques for nonparametric density deconvolution.  See Carroll \etal(2006, p.~7) for further discussion of applications to data on nutrition.\index{food|)}\index{data!nutrition data|)}

How might we correct for errors in variables?  One approach is to use methods based on deconvolution, as follows.  Let us write $Q$ for the quantity of fast food that a person admits to eating, in a food frequency questionnaire; let $Q_0$ denote the actual amount of fast food; and put $R=Q/Q_0$.  We expect that the distribution of $R$ will be skewed towards values greater than~1, and we might even have an idea of the shape of the distribution responsible for this effect, i.e.~the distribution of $\log R$.  Indeed, we typically work with the logarithm of the formula $Q=Q_0\,R$, and in that context, writing $W=\log Q$, $X=\log Q_0$ and $U=\log R$, the equation defining the variables of interest is:
\begin{equation}
W=X+U\,.\label{WXU}
\end{equation}
We have data on $W$, and from that we wish to estimate the distribution of $X$, i.e.~the distribution of the logarithm of fast-food consumption.

It can readily be seen that this problem is generally not solvable unless the distribution of $U$, and the joint distribution of $X$ and $U$, are known.  In practice we usually take $X$ and $U$ to be independent, and undertake empirical deconvolution\index{deconvolution|)} (i.e.~estimation of the distribution, or density, of $X$ from data on $W$) for several candidates for the distribution of~$U$.  If we are able to make repeated measurements of $X$, in particular to gather data on $W\j=X+U\j$ for $1\leq j\leq m$, say, then we have an opportunity to estimate the distribution of $U$ as well.

It is generally reasonable to assume that $X$, $U^{(1)}$, \ldots, $U^{(M)}$ are independent random variables. The distribution of $U$ can be estimated whenever $m\geq2$ and the distribution is uniquely determined by $|\phi_U|^2$, where $\phi_U$ denotes the \Index{characteristic function} of~$U$.  The simplest example of this type is arguably that where $U$ has a \Index{symmetric distribution} for which the characteristic function does not vanish on the real line.  One example of repeated measurements in the case $m=2$ is that where a food frequency questionnaire asks at one point how many times we visited a fast food outlet, and on a distant page, how many hamburgers or servings of fried chicken we have purchased.

The model at (\ref{WXU}) is simple and interesting, but in examples from nutrition science, and in many other problems, we generally wish to estimate the response to an \Undex{explanatory variable}, rather than the distribution of the explanatory variable.  Therefore the proper context for our food frequency questionnaire example\undex{nutrition|)} is really \Index{regression}, not distribution or density estimation.  In regression with errors in variables\index{errors in variables|)} we observe data pairs $(W,Y)$, where
\begin{equation}
W=X+U\,,\quad Y=g(X)+V\,,\label{WYfromXU}
\end{equation}
$g(x)=E(Y\mi X=x)$, and the random variable $V$, denoting an experimental error, has zero mean.  In this case the standard regression problem is altered on account of errors that are incurred when measuring the value of the explanatory variable.  In (\ref{WYfromXU}) the variables $U$, $V$ and $X$ are assumed to be independent.

The measurement error $U$, appearing in (\ref{WXU}) and (\ref{WYfromXU}), can be interpreted as the result of a `laboratory error' in determining the `dose' $X$ which is applied to the subject.  For example, a laboratory technician might use the \Index{dose} $X$ in an experiment, but in attempting to determine the dose after the experiment they might commit an error $U$, with the result that the actual dose is recorded as $X+U$ instead of~$X$.  Another way of modelling the effect of measurement error is to reverse the roles of $X$ and $W$, so that we observe $(W,Y)$ generated as
\begin{equation}
X=W+U\,,\quad Y=g(X)+V\,.\label{XYfromWU}
\end{equation}
Here a precise dose $W$ is specified, but when measuring it prior to the experiment our technician commits an error $U$, with the result that the actual dose is $W+U$.  In (\ref{XYfromWU}) it assumed that $U$, $V$ and $W$ are independent.

The measurement error model (\ref{WYfromXU}) is standard.  The alternative
model (\ref{XYfromWU}) is believed to be much less common, although in some
circumstances it is difficult to determine which of (\ref{WYfromXU}) and
(\ref{XYfromWU}) is the more appropriate.  The model at (\ref{XYfromWU}) was
first suggested by Berkson (1950)\index{Berkson, J.}, for whom it is named.

\subsection{Kernel methods}
If the measurement error $U$ were very small then we could estimate the
density $f$ of $X$, and the function $g$ in the model (\ref{WYfromXU}), using standard
kernel methods.  For example, given data $X_1$, \ldots, $X_n$ on $X$ we could take
\begin{equation}
\hf(x)=\oonh\,\sumion K\Big(\frac{x-X_i}h\Big)\label{fhat}
\end{equation}
to be our estimator of $f(x)$.  Here $K$ is a kernel function and $h$, a
positive quantity, is a bandwidth\index{bandwidth|(}.  Likewise, given data $(X_1,Y_1)$, \ldots, $(X_n,Y_n)$ on $(X,Y)$ we could take
\begin{equation}
\hg(x)=\frac{\sumi Y_i\,K\{(x-X_i)/h\}}{\sumi K\{(x-X_i)/h\}}\label{ghat}
\end{equation}
to be our estimator of $g(x)$, where $g$ is as in the model at~(\ref{WYfromXU}).

The estimator at (\ref{fhat}) is a standard kernel density
estimator\index{kernel methods!kernel density estimator}, and is itself a
probability density if we take $K$ to be a density.  It is consistent under
particularly weak conditions, for example if $f$ is continuous and $h\ra0$ and
$nh\rai$ as $n$ increases.  Density estimation\index{density estimation} is discussed at length by
Silverman (1986)\index{Silverman, B. W.} and Scott~(1992)\index{Scott, D. W.}.
The estimator $\hg$, which we generally also compute by taking $K$ to be a
probability density, is often referred to as the `local
constant'\index{density estimation!local
constant estimator} or Nadaraya--Watson\index{Nadaraya, E. A.!Nadaraya--Watson estimator} estimator of~$g$.  The first of these names follows from the fact that $\hg(x)$ is the result of fitting a constant to the data by \Index{local least squares}:
\begin{equation}
\hg(x)=\argmin_c\,\sumion(Y_i-c)^2\,K\Big(\frac{x-X_i}h\Big)\,.\label{ghatArgmin}
\end{equation}
The estimator $\hg$ is also consistent under mild conditions, for example if
the variance of the error, $V$, in (\ref{WYfromXU}) is finite, if $f$ and $g$
are continuous,  if $f>0$ at the point $x$ where we wish to estimate~$g$, and
if $h\to 0$ and $nh\to\infty$ as $n$ increases.  General kernel methods are
discussed by Wand and Jones (1995)\index{Wand, M. P.}\index{Jones, M. C.},
and \Index{statistical smoothing} is addressed by
Simonoff~(1996)\index{Simonoff, J. S.}.

Local constant estimators have the advantage of being relatively robust
against uneven spacings in the sequence $X_1$, \ldots, $X_n$.  For example,
the ratio at (\ref{ghat}) never equals a nonzero number divided by zero.
However, local constant estimators are particularly susceptible
to \Index{boundary bias}.  In particular, if the density of $X$ is supported
and bounded away from zero on a compact interval, then $\hg$, defined by
(\ref{ghat}) or (\ref{ghatArgmin}), is generally inconsistent at the endpoints
of that interval.  Issues of this type have motivated the use of local
polynomial estimators\index{density estimation!local polynomial estimator}, which are defined by $\hg(x)=\hatc_0(x)$ where, in a generalisation of~(\ref{ghatArgmin}),
\begin{equation}
(\hatc_0(x),\ldots,\hatc_p(x))
=\argmin_{(c_0,\ldots,c_p)}\,\sumion\bigg\{Y_i-\sum_{j=0}^p\,c_j\,(x-X_i)^j\bigg\}^2\,
K\Big(\frac{x-X_i}h\Big)\,.\label{chat}
\end{equation}
See, for example, Fan and Gijbels~(1996)\index{Fan, J.}\index{Gijbels, I.}.  In (\ref{chat}), $p$ denotes the degree of the locally fitted polynomial.  The estimator $\hg(x)=\hatc_0(x)$, defined by (\ref{chat}), is also consistent under the conditions given earlier for the estimator defined by (\ref{ghat}) and~(\ref{ghatArgmin}).

In the particular case $p=1$ we obtain a local-linear
estimator\index{density estimation!local linear estimator|(} of~$g(x)$:
\begin{equation}
\hg(x)=\frac{S_2(x)\,T_0(x)-S_1(x)\,T_1(x)}{S_0(x)\,S_2(x)-S_1(x)^2}\,,\label{loclin}
\end{equation}
where
\begin{equation}
\begin{split}
S_r(x)&=\oonh\,\sumion
\bigg(\frac{x-X_i}h\bigg)^{\!r}\>
K\bigg(\frac{x-X_i}h\bigg)\,,\\
T_r(x)&=\oonh\,\sumion Y_i\,
\bigg(\frac{x-X_i}h\bigg)^{\!r}\>
K\bigg(\frac{x-X_i}h\bigg)\,,
\end{split}\label{SxTx}
\end{equation}
$h$ denotes a bandwidth\index{bandwidth|)} and $K$ is a kernel function.

Estimators of all these types can be quickly extended to cases where errors in variables are present, for example as in the models at (\ref{WXU}) and (\ref{WYfromXU}), simply by altering the kernel function $K$ so that it acts to cancel out the influence of the errors.  We shall give details in section~2.  Section~3 will discuss recently introduced methodology which, from some viewpoints looks quite different from, but is actually almost identical to, kernel methods.

\section{Methodology and theory}
\subsection{Definitions of estimators}
We first discuss a generalisation of the estimator at (\ref{fhat}) to the case where
there are errors in the observations of $X_i$, as per the model at~(\ref{WXU}).  In
particular, we assume that we observe data $W_1$, \ldots, $W_n$ which are
independent and identically distributed as $W=X+U$, where $X$ and $U$ are
independent and the distribution of $U$ has known \Index{characteristic
function} $\phi_U$ which does not vanish anywhere on the real line.  Let $K$
be a kernel function, write $\phi_K=\int e^{itx}\,K(x)\,dx$ for the associated
Fourier transform\index{Fourier, J. B. J.!Fourier transform|(}, and define
\begin{equation}
K_U(x)=\ootp\int e^{-itx}\;\frac{\phi_K(t)}{\phi_U(t/h)}\;dt\,.\label{KU}
\end{equation}
Then, to construct an estimator $\hf$ of the density $f=f_X$ of $X$, when all
we observe are the contaminated data $W_1$, \ldots, $W_n$, we simply replace $K$ by $K_U$, and $X_i$ by $W_i$, in the definition of $\hf$ at (\ref{fhat}), obtaining the estimator
\begin{equation}
\hf\decon(x)=\oonh\,\sumion K_U\Big(\frac{x-W_i}h\Big)\,.\label{fhatdecon}
\end{equation}
Here the subscript `decon' signifies that $\hf\decon$ involves empirical
deconvolution\index{deconvolution|(}.  The adjustment to the kernel takes care of the measurement
error\index{measurement error|)}, and results in consistency in a wide variety of settings.  Likewise, if
data pairs $(W_1,Y_1)$, \ldots, $(W_n,Y_n)$ are generated under the model at
(\ref{WYfromXU}) then, to construct the local constant estimator\index{density
estimation!local constant estimator} at (\ref{ghat}), or the local linear
estimator\index{density estimation!local linear estimator|)} defined by (\ref{loclin}) and (\ref{SxTx}), all we do is replace each $X_i$ by $W_i$, and $K$ by~$K_U$.  Other local polynomial estimators\index{density estimation!local polynomial estimator} can be calculated using a similar rule, replacing $h^{-r}(x-X_i)^rK\{(x-X_i)/h\}$ in $S_r$ and $T_r$ by $K_{U,r}\{(x-W_i)/h\}$, where
$$
K_{U,r}(x)=\frac1{2\pi i^r}\int e^{-itx}\;\frac{\phi_K^{(r)}(t)}{\phi_U(t/h)}\;dt\,.
$$

The estimator at (\ref{fhatdecon}) dates from work of Carroll and Hall
(1988)\index{Carroll, R. J.} and Stefanski and Carroll
(1990)\index{Stefanski, L.}.  Deconvolution-kernel regression
estimators\index{kernel methods!deconvolution-kernel estimator} in the
local-constant\index{density estimation!local constant estimator} case were developed by Fan and
Truong (1993)\index{Fan, J.}\index{Truong, Y. K.}, and extended to the general
local polynomial\index{density estimation!local polynomial estimator} setting by Delaigle \etal(2009).

The kernel $K_U$ is deliberately constructed to be the function whose Fourier
transform\index{Fourier, J. B. J.!Fourier transform} is $\phi_K/\phi_U$.  This
adjustment permits cancellation of the influence of errors in variables, as
discussed at the end of section~1.3.  To simplify calculations, for example
computation of the integral in (\ref{WYfromXU}), we generally choose $K$ not
to be a density function but to be a smooth, symmetric function for which
$\phi_K$ vanishes outside a compact interval.  The commonly-used candidates
for $\phi_K$ are proportional to functions that are used for $K$, rather than
$\phi_K$, in the case of regular kernel estimation discussed in section~1.3.
For example, kernels $K$ for which $\phi_K(t)=(1-|t|^r)^s$ for $|t|\leq1$, and
$\phi_K(t)=0$ otherwise, are common; here $r$ and $s$ are integers.  Taking
$r=2s=2$, $r=s=2$ and $r=\frac23\,s=2$ corresponds to the Fourier
inverses\index{Fourier, J. B. J.!Fourier inverse} of the biweight, quartic and
triweight kernels, respectively.  Taking $s=0$ gives the inverse of the
uniform kernel, i.e.~the \index{kernel methods!sinc kernel}sinc kernel, which we shall meet again in section~3. Further information about kernel choice is given by Delaigle and Hall (2006).

These kernels, and others, have the property that $\phi_K(t)=1$ when $t=0$,
thereby guaranteeing that $\int K=1$.  The latter condition ensures that the
density estimator\index{density estimation}, defined at (\ref{fhatdecon}) and
constructed using this kernel, integrates to~1.  (However, the estimator
defined by (\ref{fhatdecon}) will generally take negative values at some
points~$x$.)  The normalisation property is not so important when the kernel
is used to construct regression estimators\index{regression!regression estimator}, where the effects of multiplying $K$ by a constant factor cancel from the `deconvolution' versions of formulae (\ref{ghat}) and~(\ref{loclin}), and likewise vanish for all deconvolution-kernel estimators\index{kernel methods!deconvolution-kernel estimator} based on local polynomial methods\index{density estimation!local polynomial estimator}.

Note that, as long as $\phi_K$ and $\phi_U$ are supported either on the whole real line or on a symmetric compact domain, the kernel $K_U$, defined by (\ref{KU}), and its generalised form $K_{U,r}$, are real-valued. Indeed, using properties of the complex conjugate of Fourier transforms\index{Fourier, J. B. J.!Fourier transform|)} of real-valued functions, and the change of variable $u=-t$, we have, using the notation $\overline a(t)$ for the complex conjugate of a complex-valued function $a$ of a real variable $t$,
\begin{align*}
\overline K_{U,r}(x)
&=(-1)^{-r} \frac1{2\pi i^r}\int e^{itx}\;\frac{\overline{\phi_K^{(r)}}(t)}{\overline\phi_U(t/h)}\;dt\\
&=(-1)^{-r} \frac1{2\pi i^r}\int e^{itx}\;\frac{(-1)^{-r}\phi_K^{(r)}(-t)}{\phi_U(-t/h)}\;dt\\
&= \frac1{2\pi i^r}\int e^{-iux}\;\frac{\phi_K^{(r)}(u)}{\phi_U(u/h)}\;du
=K_{U,r}(x).
\end{align*}
In practice it is almost always the case that the distribution of $U$ is
symmetric\index{symmetric distribution}, and in the discussion of variance in section~2.2, below, we shall make this assumption.  We shall also suppose that $K$ is symmetric, again a condition which holds almost invariably in practice.

The estimators discussed above were based on the assumption that
the characteristic function\index{characteristic function|(} $\phi_U$ of the errors in variables is
known.  This enabled us to compute the deconvolution kernel $K_U$
at~(\ref{KU}).  In cases where the distribution of $U$ is not known, but can
be estimated from replicated data (see section~1.2), we can replace $\phi_U$
by an estimator of it and, perhaps after a little regularisation, compute an
empirical version of~$K_U$.  This can give good results, in both theory and
practice.  In particular, in many cases the resulting estimator of the density
of $X$, or the regression mean\index{regression!regression mean} $g$, can be shown to have the same first-order properties as estimators computed under the assumption that the distribution of $U$ is known.   Details are given by Delaigle \etal(2008).

Methods for choosing the \Index{smoothing parameter}, $h$, in the estimators
discussed above have been proposed by Hesse (1999)\index{Hesse, C.}, Delaigle
and Gijbels (2004a,b) and Delaigle and Hall~(2008).

\subsection{Bias and variance}\index{bias}
The expected value of the estimator at (\ref{fhatdecon}) equals
\begin{align}
E\{\hf\decon(x)\}&=\frac1{2\pi h}\int
  E\big[e^{-it\{x-W\}/h}\big]\;\frac{\phi_K(t)}{\phi_U(t/h)}\;dt\notag \\
&=\frac1{2\pi}\int
  e^{-itx}\frac{\phi_K(ht)}{\phi_U(t)}\;\phi_X(t)\,\phi_U(t)\,dt\notag \\
&=\frac1{2\pi}\int e^{-itx}\phi_K(ht)\,\phi_X(t)\,dt
=\frac1h\int K(u/h)\,f(x-u)\,du\notag \\
&=E\{\hf(x)\}\,,\label{Efhatdecon}
\end{align}
where the first equality uses the definition of $K_U$, and the fourth equality
uses Plancherel's identity\index{Plancherel, M.!Plancherel's identity}.  Therefore the deconvolution estimator $\hf\decon(x)$, calculated from data contaminated by measurement errors, has exactly the same mean, and therefore the same bias, as $\hf(x)$, which would be computed using values of $X_i$ observed without measurement error.  This confirms that using the deconvolution kernel estimator does indeed allow for cancellation of measurement errors, at least in terms of their presence in the mean.

Of course, variance is a different matter.  Since $\hf\decon(x)$ equals a sum of independent random variables then
\begin{align}
&\var\{\hf\decon(x)\}\notag \\
&{}\qquad=\big(nh^2\big)\mo\,\var\Big\{K_U\Big(\frac{x-W}h\Big)\Big\}\notag \\
&{}\qquad\sim(nh)\mo\,f_W(x)\,\int K_U^2
=\frac{f_W(x)}{2\pi nh}\;\int\phi_K(t)^2\,|\phi_U(t/h)|\mt\,dt\,.\label{varfhat}
\end{align}
(Here the relation $\sim$ means that the ratio of the left- and right-hand
sides converges to~1 as $h\ra0$.)  Thus it can be seen that the variance of
$\hf\decon(x)$ depends intimately on tail behaviour\index{tail behaviour|(} of the characteristic function\index{characteristic function|)} $\phi_U$ of the measurement-error distribution.

If $\phi_K$ vanishes outside a compact set, which, as we noted in section~2.1,
is generally the case, and if $|\phi_U|$ is asymptotic to a positive regularly
varying function\index{regular variation} $\psi$ (see Bingham \etalc
1989)\index{Bingham, N. H.}\index{Goldie, C. M.}\index{Teugels, J. L.}, in the sense that $|\phi_U(t)|\asymp\psi(t)$ (meaning that the ratio of both sides is bounded away from zero and infinity as $t\rai$), then the integral on the right-hand side of (\ref{Efhatdecon}) is bounded between two constant multiples of $\psi(1/h)\mt$ as $h\ra0$.  Therefore by (\ref{varfhat}), provided that $f_W(x)>0$,
\begin{equation}
\var\{\hf\decon(x)\}\asymp(nh)\mo\,\psi(1/h)\mt\label{varasymp}
\end{equation}
as $n$ increases and $h$ decreases.
Recall that we are assuming that $f_U$ and $K$ are both symmetric functions.

If the density $f$ of $X$ has two bounded and continuous derivatives, and if $K$ is bounded and symmetric and satisfies $\int x^2\,|K(x)|\,dx<\infty$, then the bias of $\hf\decon$ can be found from (\ref{Efhatdecon}), using elementary calculus and arguments familiar in the case of standard kernel estimators:
\begin{align}
\bias(x)&=E\{\hf\decon(x)\}-f(x)=E\{\hf(x)\}-f(x)\notag \\
&=\int K(u)\,\{f(x-hu)-f(x)\}\,du
=\thf\,h^2\,\ka\,f''(x)+o\big(h^2\big)\label{bias}
\end{align}
as $h\ra0$, where $\ka=\int x^2\,K(x)\,dx$. Therefore, provided that $f''(x)\neq0$, the bias of the conventional kernel estimator $\hf(x)$ is exactly of size $h^2$ as $h\ra0$.  Combining this property, (\ref{Efhatdecon}) and (\ref{varasymp}) we deduce a relatively concise asymptotic formula for the \Index{mean squared error} of~$\hf\decon(x)$:
\begin{equation}
E\{\hf\decon(x)-f(x)\}^2\asymp h^4+(nh)\mo\,\psi(1/h)\mt\,.\label{MSE}
\end{equation}
For a given error distribution we can work out the behaviour of $\psi(1/h)$ as
$h\ra0$, and then from (\ref{MSE}) we can calculate the
optimal \Index{bandwidth} and determine the exact rate of convergence of
$\hf\decon(x)$ to $f(x)$, in mean square.  In many instances this rate is
optimal, in a \Index{minimax} sense; see, for example, Fan~(1991)\index{Fan,
J.}.  It is also generally optimal in the case of the
errors-in-variables\index{errors in variables} regression estimators discussed in section~2.1, based on deconvolution-kernel versions of local polynomial estimators\index{density estimation!local polynomial estimator}.  See Fan and Truong~(1993)\index{Fan, J.}\index{Truong, Y. K.}.

Therefore, despite their almost naive simplicity, deconvolution-kernel
estimators of densities and regression functions have features that can hardly
be bettered by more complex, alternative approaches.  The results derived in
the previous paragraph, and their counterparts in the regression case, imply
that the estimators are limited by the extent to which they can
recover \index{high-frequency information} from the data.  (This is reflected
in the fact that the rate of decay of the tails\index{tail behaviour|)} of $\phi_U$ drives the results on convergence rates.)  However, the fact that the estimators are nevertheless optimal, in terms of their rates of convergence, implies that this restriction is inherent to the problem, not just to the estimators; no other estimators would have a better convergence rate, at least not uniformly in a class of problems.

\section{Relationship to minimum contrast methods}\index{minimum contrast methods|(}
\subsection{Deconvolution kernel estimators based on the sinc kernel}
The sinc, or Fourier integral, kernel\index{kernel methods!sinc kernel|(} is given by
\begin{equation}
L(x)=\begin{cases}(\pi x)\mo\,\sin(\pi x)&\text{if $x\neq0$}\\
                                  1&\text{if $x=0\,.$}\end{cases}\label{Lx}
\end{equation}
Its Fourier transform\index{Fourier, J. B. J.!Fourier transform}, defined as a
Riemann integral\index{Riemann, G. F. B.!Riemann integral}, is the `boxcar
function'\index{boxcar function}, $\phi_L(t)=1$ if $|t|\leq1$ and $\phi_L(t)=0$ otherwise.  In particular, $\phi_L$ vanishes outside a compact set, which property, as we noted in section~2.1, aids computation.  The version of $K_U$, at (\ref{KU}), for the sinc kernel is
$$
L_U(x)=\ootp\int_{-1}^1 e^{-itx}\,\phi_U(t/h)\mo\,dt
=\frac1\pi\int_0^1\cos(tx)\,\phi_U(t/h)\mo\,dt\,,
$$
where the second identity holds if the distribution of $U$ is
symmetric\index{symmetric distribution} and has no zeros on the real line.

The kernel $L$ is sometimes said to be of `infinite order'\index{kernel
methods!infinite order kernel}, in the sense that if $a$ is any function with an infinite number of bounded, integrable derivatives then
\begin{equation}
\int\bigg[\int\{a(x+hu)-a(x)\}\,L(u)\,du\bigg]^2\,dx=O\big(h^r\big)\label{intintaa}
\end{equation}
as $h\da0$, for all $r>0$.  If $K$ were of finite order then (\ref{intintaa}) would hold only for a finite range of values of $r$, no matter how many derivatives the function $a$ enjoyed.  For example, if $K$ were a symmetric function for which $\int u^2\,K(u)\,du\neq0$, and if we were to replace $L$ in (\ref{intintaa}) by $K$, then (\ref{intintaa}) would hold only for $r\leq4$, not for all~$r$.  In this case we would say that $K$ was of second order\index{kernel
methods!second-order kernel|(}, because
$$
\int\{a(x+hu)-a(x)\}\,K(u)\,du=O\big(h^2\big)\,.
$$

If we take $a$ to be the density, $f$, of the random variable $X$, and take $K$ in the definition of $\hf$ at (\ref{fhat}) to be the sinc kernel, $L$, then (\ref{intintaa}) equals the integral of the squared \Index{bias} of~$\hf$.  Therefore, in the case of a very smooth density, the `infinite order' property of the sinc kernel ensures particularly small bias, in an average sense.

Properties of conventional kernel density estimators, but founded on the sinc
kernel, for data without measurement errors, have been studied by, for
example, Davis (1975, 1977)\index{Davis, K. B.}.  Glad \etal(1999)\index{Glad,
I. K.}\index{Hjort, N. L.}\index{Ushakov, N.}have provided a good survey of
properties of sinc kernel methods for density estimation, and have argued that
those estimators have received an unfairly bad press.  Despite criticism of
sinc kernel estimators (see e.g.~Politis and Romano,~1999)\index{Politis,
D. N.}\index{Romano, J. P.}, the approach is ``more accurate for quite moderate values of the sample size, has better asymptotics in non-smooth cases (the density to be estimated has only first derivative), [and] is more convenient for bandwidth selection etc''\index{bandwidth} than its conventional competitors, suggest Glad \etal(1999).

The property of greater accuracy is borne out in both theoretical and numerical studies, and derives from the infinite-order property noted above.  Indeed, if $f$ is very smooth then the low level of average squared bias can be exploited to produce an estimator $\hf$ with particularly low mean squared error, in fact of order $n\mo$ in some cases.  The most easily seen disadvantage of sinc-kernel density estimators is their tendency to suffer from spurious oscillations, inherited from the infinite number of oscillations of the kernel itself.

These properties can be expected to carry over to density and regression
estimators based on
\index{data!contaminated data}contaminated data, when we use the sinc kernel.  To give a
little detail in the case of density estimation from data contaminated by
measurement errors, we note that if the density $f$ of $X$ is infinitely
differentiable, but we observe only the contaminated data $W_1$, \ldots, $W_n$
distributed as $W$, generated as at (\ref{WXU}); if we use the density estimator at
(\ref{fhat}), but computed using $K=L$, the sinc kernel; and if $|\phi_U(t)|\geq
C\,(1+|t|)^{-\a}$ for constants $C$, $\a>0$; then, in view of (\ref{Efhatdecon}), (\ref{varfhat}) and (\ref{intintaa}), we have for all $r>0$,
\begin{align}
&\int\{\hf\decon(x)-f(x)\}^2\,dx\notag \\
&{}\qquad=\int\{E\hf(x)-f(x)\}^2
+\big(nh^2\big)\mo\,\int\var\Big\{L_U\Big(\frac{x-W}h\Big)\Big\}\,dx\notag \\
&{}\qquad\leq\int\bigg[\int\{f(x+hu)-f(x)\}\,L(u)\,du\bigg]^2\,dx
+(nh)\mo\,\int L_U^2\notag \\
&{}\qquad=O\bigg\{h^r+(nh)\mo\,\int_{-1}^1|\phi_U(t/h)|\mt\,dt\bigg\}\notag \\
&{}\qquad=O\Big\{h^r+\big(nh^{2\a+1}\big)\mo\Big\}\,.\label{intfhat}
\end{align}
It follows that, if $f$ has infinitely many integrable derivatives and if the tails of $\phi_U(t)$ decrease at no faster than a polynomial rate as $|t|\rai$, then the \Index{bandwidth} $h$ can be chosen so that the mean integrated squared error of a deconvolution kernel estimator of $f$, using the sinc kernel, converges at rate $O(n^{\ep-1})$ for any given $\ep>0$.

This very fast rate of convergence contrasts with that which occurs if the
kernel $K$ is of only finite order.  For example, if $K$ is a second-order
kernel, in which case (\ref{intintaa}) holds only for $r\leq 4$ when $L$ is replaced by $K$, the argument at (\ref{intfhat}) gives:
$$
\int\{\hf\decon(x)-f(x)\}^2\,dx
=O\Big\{h^4+\big(nh^{2\a+1}\big)\mo\Big\}\,.
$$
The fastest rate of convergence of the right-hand side to zero is attained with $h=n^{-1/(2\a+5)}$, giving
$$
\int\{\hf\decon(x)-f(x)\}^2\,dx=O\big(n^{-4/(2\a+5)}\big)\,.
$$
In fact, this is generally the best rate of convergence of mean integrated
squared error that can be obtained using a second-order kernel when
the \Index{characteristic function} $\phi_U$ decreases like $|t|^{-\a}$ in the
tails\index{tail behaviour}, even if the density $f$ is exceptionally smooth.  Nevertheless, second-order kernels\index{kernel methods!second-order kernel|)} are often preferred to the sinc kernel in practice, since they do not suffer from the unwanted oscillations that afflict estimators based on the sinc kernel.\index{kernel methods!sinc kernel|)}

\subsection{Minimum contrast estimators, and their relationship to
deconvolution kernel estimators}
In the context of the measurement error model at (\ref{WXU}), Comte {\em et
al.} (2007)\index{Comte, F.|(}\index{Rozenholc, Y.|(}\index{Taupin, M.-L.|(} suggested an interesting minimum contrast estimator of the density $f$ of $X$.  Their approach has applications in a variety of other settings (see Comte \etalc2006, 2008; Comte and Taupin, 2007), including to the regression model at (\ref{WYfromXU}), and the conclusions we shall draw below apply in these cases too.  Therefore, for the sake of brevity we shall treat only the density deconvolution problem.

To describe the minimum contrast estimator in that setting, define
$$
\ha\kl=\frac1{2\pi n}\,
\sumjon\int\exp(it\,W_j)\,\phi_{L\kl}(t)\,\phi_U(t)^{-1}\,dt\,,
$$
where $\phi_{L\kl}$ denotes the Fourier transform\index{Fourier, J. B. J.!Fourier transform} of the function $L\kl$ defined by $L\kl(x)=\ell\half\,L(\ell\,x-k)$, $k$ is an integer and $\ell>0$.  In this notation the minimum contrast nonparametric density estimator~is
$$
\tf(x)=\sum_{k=-k_0}^{k_0}\,\ha\kl\,L\kl(x)\,.
$$
There are two tuning parameters, $k_0$ and~$\ell$.  Comte \etal(2007) suggest choosing $\ell$ to minimise a penalisation criterion.

The resulting minimum contrast estimator is called a penalised contrast
density estimator\index{penalised contrast methods}. The penalisation
criterion suggested by Comte \etal(2007) for choosing $\ell$ is related to
cross-validation, although its exact form, which involves the choice of
additional terms and multiplicative constants, is based on \Index{simulation}
experiments.  It is clear on inspecting the definition of $\tf$ that $\ell$
plays a role similar to that of the inverse of \Index{bandwidth} in a
conventional deconvolution kernel estimator\index{kernel methods!deconvolution-kernel estimator}.  In particular, $\ell$ should diverge to infinity with~$n$.  Comte \etal(2007) suggest taking $k_0=2^m-1$, where $m\geq\log_2(n+1)$ is an integer.  In numerical experiments they use $m=8$, which gives good performance in the cases they consider.  More generally, $k_0/\ell$ should diverge to infinity as sample size increases.

The minimum contrast density estimator of Comte \etal(2007) is actually very
close to the standard deconvolution kernel density estimator at (\ref{fhat}),
where in the latter we use the sinc kernel\index{kernel methods!sinc kernel} at~(\ref{Lx}).  Indeed, as the theorem below shows, the two estimators are exactly equal on a grid, which becomes finer as the bandwidth, $h$, for the sinc kernel density estimator decreases.  However, this relationship holds only for values of $x$ for which $|x|\leq k_0/\ell$; for larger values of $|x|$ on the grid, $\tf(x)$ vanishes.  (This property is one of the manifestations of the fact that, as noted earlier, $k$ and $\ell$ generally should be chosen to depend on sample size in such a manner that $k_0/\ell\rai$ as $n\rai$.)

\begin{nthm}Let $\hf\decon$ denote the deconvolution kernel density estimator at $(\ref{fhat})$, constructed using the sinc kernel and employing the bandwidth $h=\ell\mo$.  Then, for any point $x=hk$ with $k$ an integer, we have
$$
\tf(x)=\begin{cases}\hf\decon(x)&\text{if $|x|\leq k_0/\ell$}\\
                         0&\text{if $|x|>k_0/\ell\,.$}\end{cases}
$$
\end{nthm}\index{bandwidth}

A proof of the theorem will be given in section~3.3.  Between grid points the estimator $\tf$ is a nonstandard interpolation of values of the kernel estimator $\hf\decon$.  Note that, if we take $h=\ell\mo$, the weights $L(\ell x-k)=L\{(x-hk)/h\}$ used in the interpolation decrease quickly as $k$ moves further from $x/h$, and, except for small $k$, neighbour weights are close in magnitude but differ in sign. (Here $L$ is the sinc kernel\index{kernel methods!sinc kernel} defined at~(\ref{Lx}).)  In effect, the interpolation is based on rather few values $\hf\decon(k/\ell )$ corresponding to those $k$ for which $k$ is close to $x/h$.

In practice the two estimators are almost indistinguishable. For example,
Figure~3.1 compares them using the \Index{bandwidth} that minimises the
integrated squared difference between the true density and the estimator, for
one generated sample in the case where $X$ is normal N$(0,1)$, $U$ is
Laplace\index{Laplace, P.-S.!Laplace distribution} with $\var(U)/\var(X)=0.1$, and $n=100$ or $n=1000$. In the left graphs the two estimators can hardly be distinguished.  The right graphs show magnifications of these estimators for $x\in[-\thf,0]$.  Here it can be seen more clearly that the minimum contrast estimator is an approximation of the deconvolution kernel estimator, and is exactly equal to the latter at $x=0$.

\begin{figure}[h!t]
\mbox{\includegraphics[scale=0.4]{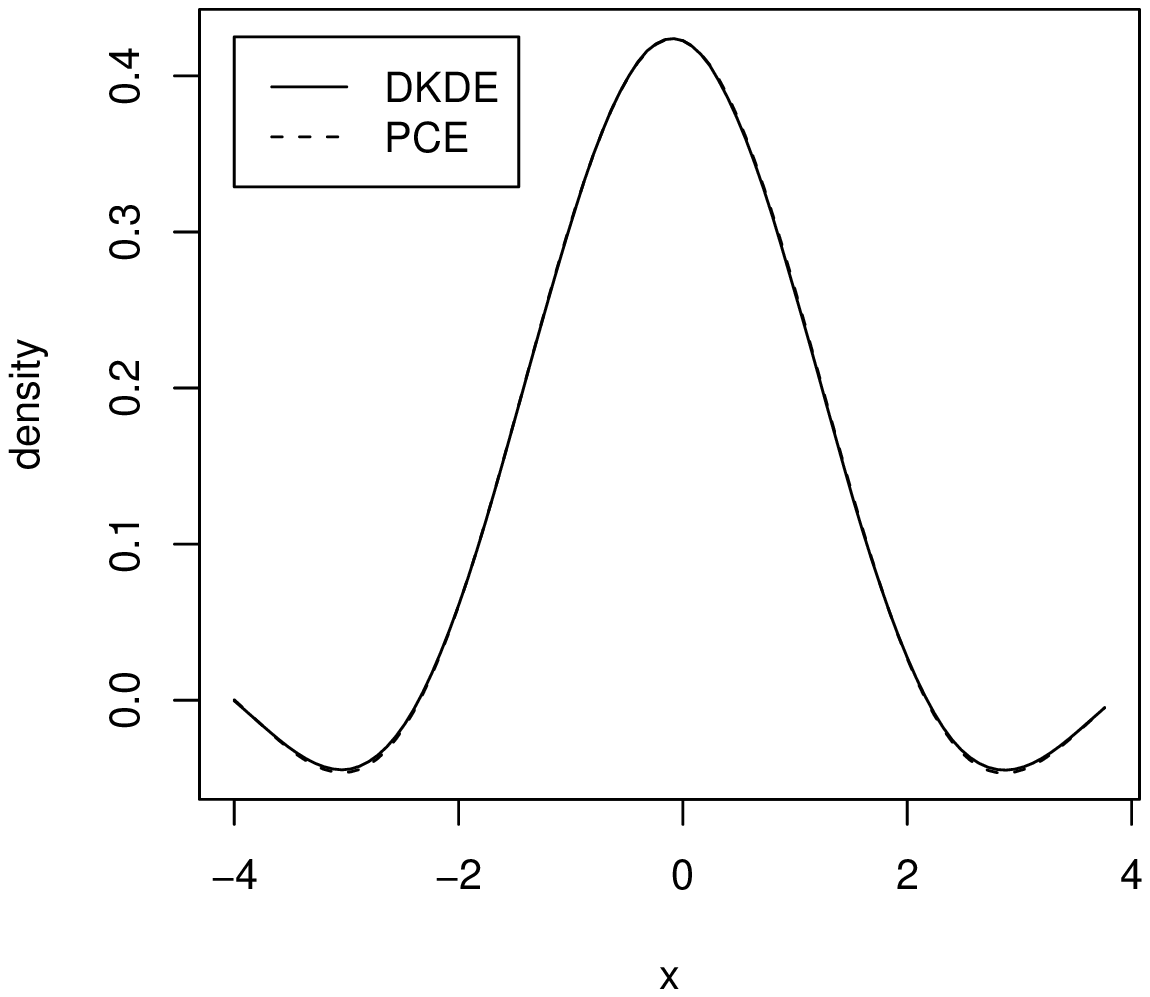}\quad
\includegraphics[scale=0.4]{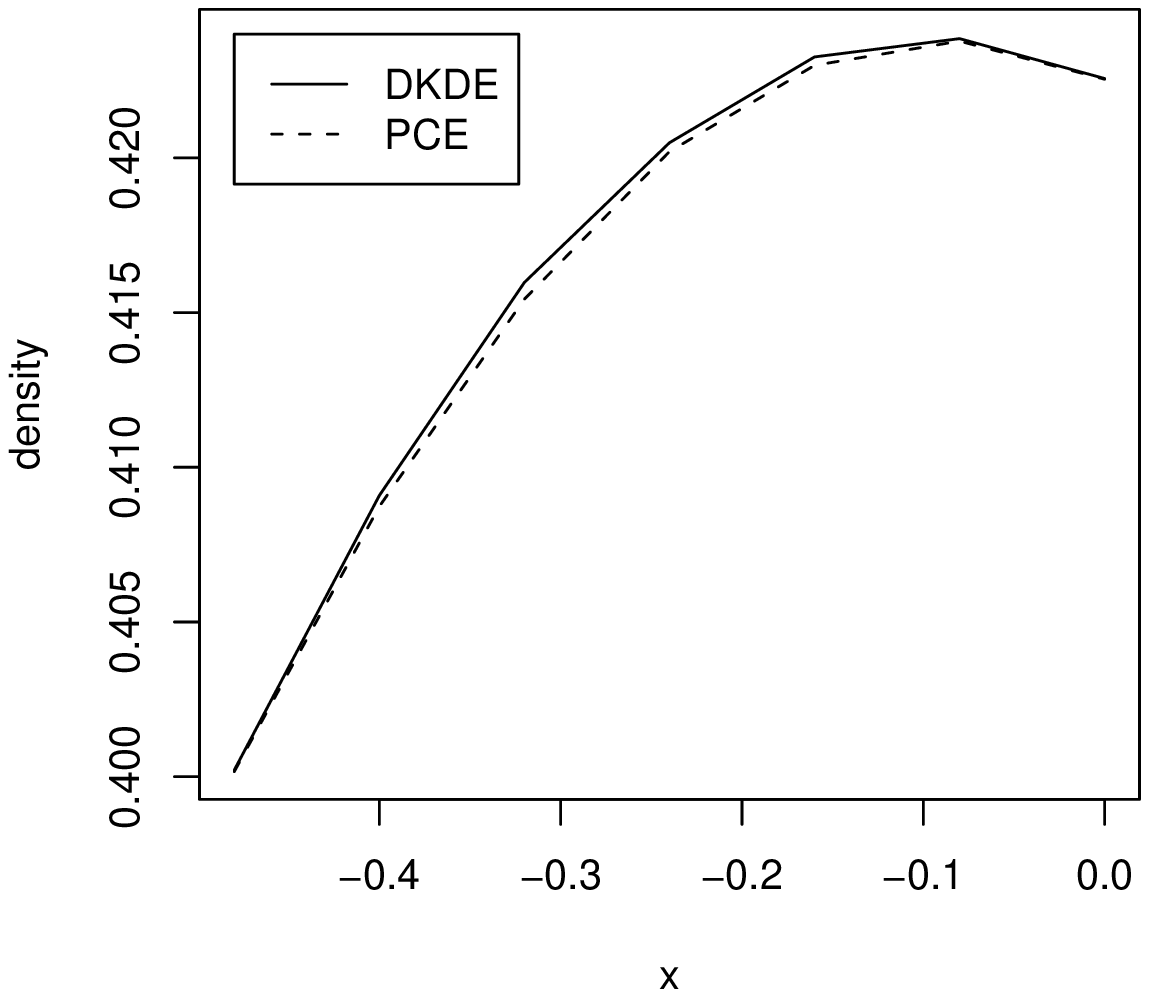}}

\mbox{\includegraphics[scale=0.4]{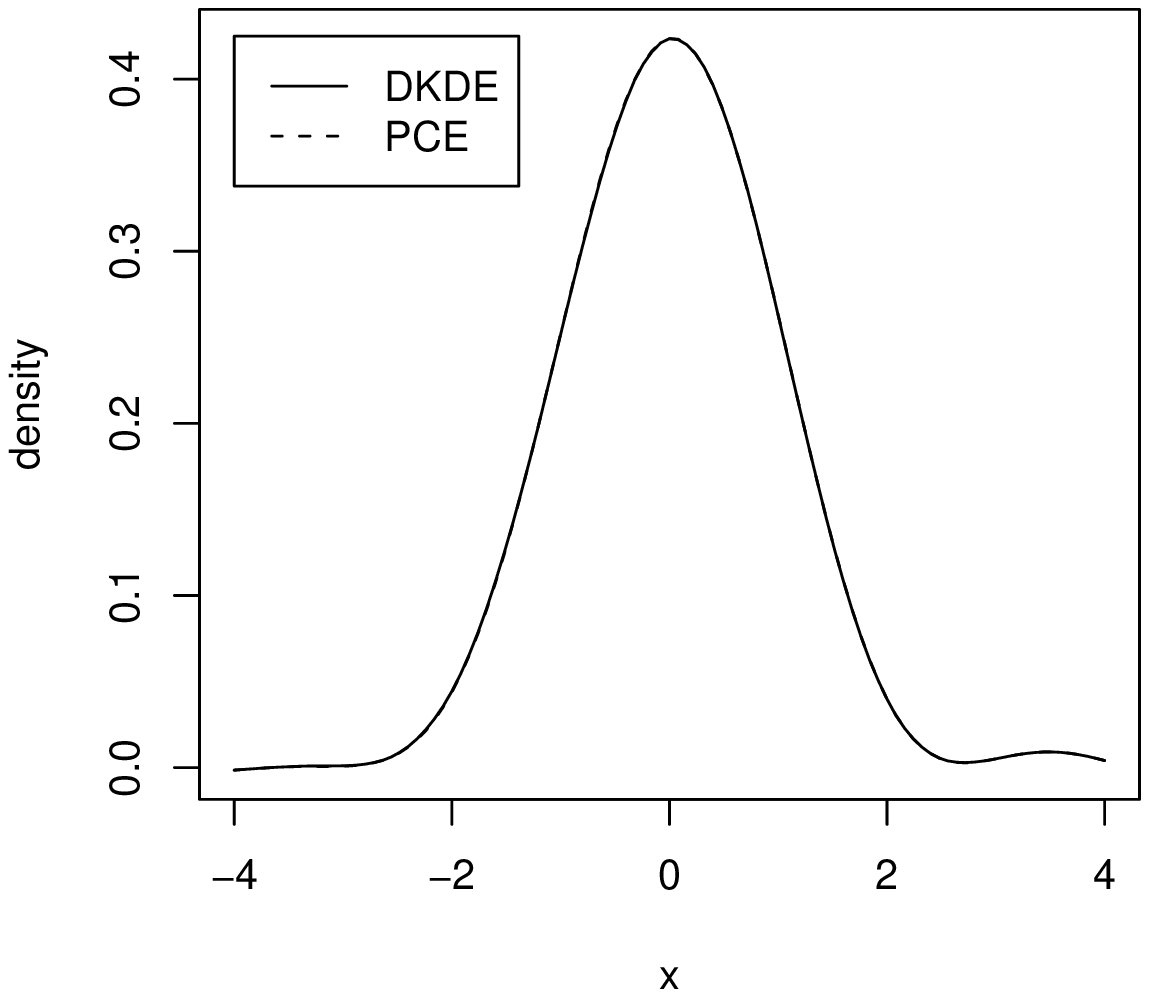}\quad
\includegraphics[scale=0.4]{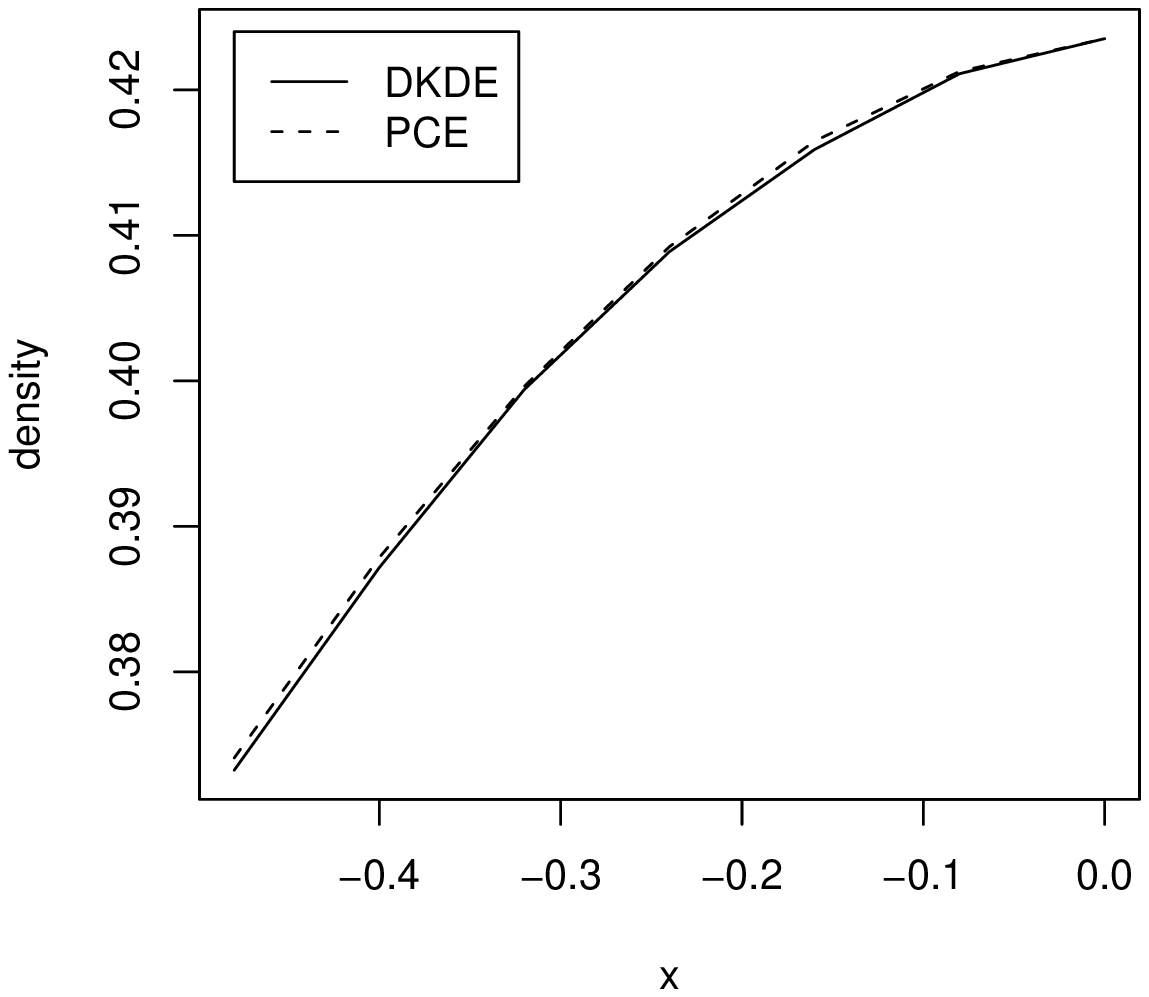}}
\caption{Deconvolution kernel density estimator (DKDE) and minimum contrast
  estimator (PCE) for a particular sample of size $n=100$ (upper panels) or
  $n=1000$ (lower panels) in the case $\var(U)/\var(X)=0.1$. Right panels show
  magnifications of the estimates for $x\in[-0.5,0]$ in the respective upper
  panels.}
\end{figure}

These results highlight the fact that the differences in performance between
the two estimators derive more from different tuning parameter choices than
from anything else.  In their comparison, Comte \etal(2007)\index{Comte,
F.|)}\index{Rozenholc, Y.|)}\index{Taupin, M.-L.|)} used a minimum contrast
estimator with the sinc kernel $L$ and a \Index{bandwidth} chosen by
penalisation, whereas for the deconvolution kernel estimator they employed a
conventional second-order kernel $K$ and a different bandwidth-choice
procedure.  Against the background of the theoretical analysis in section~3.1,
the different kernel choices (and different ways of choosing smoothing
parameters)\index{smoothing
parameter} explain the differences observed between the penalised contrast
density estimator\index{penalised contrast methods} and the deconvolution
kernel density estimator based on a second-order kernel\index{kernel
methods!second-order kernel}.

\subsection{Proof of Theorem}
Note that $\phi_{L\kl}(t)=\ell\mhf\,\exp(itk/\ell)\,\phi_L(t/\ell)$ and
$$
\ha\kl=\frac1{2n\pi\ell\half}\;
\sumjon\int_{-\ell\pi}^{\ell\pi}\exp\big\{-it\,\big(k\,\ell\mo-W_j\big)\big\}\,
\frac{\phi_L(t/\ell)}{\pi_U(t)}\;dt\,.
$$
Therefore,
\begin{align}
&\tf(x)\notag \\
&{}\quad=\frac1{2n\pi}\sum_{k=-k_0}^{k_0}L(\ell x-k)
\sumjon\int_{-\ell\pi}^{\ell\pi}\exp\big\{-it\big(k\ell\mo-W_j\big)\big\}
\frac{\phi_L(t/\ell)}{\pi_U(t)}\,dt\notag \\
&{}\quad=\sum_{k=-k_0}^{k_0}\,L(\ell x-k)\,\hf\decon(k/\ell)\,.\label{tildef}
\end{align}
If $r$ is a nonzero integer then $L(r)=0$.  Therefore, if $x=kh=s/\ell$ for an integer $s$ then $L(\ell x-k)=0$ whenever $k\neq s$, and $L(\ell x-k)=1$ if $k=s$.  Hence, (\ref{tildef}) implies that $\tf(x)=\hf\decon(x)$ if $|k|\leq k_0$, and $\tf(x)=0$ otherwise.\index{kernel methods|)}\index{deconvolution|)}\index{minimum contrast methods|)}

\end{document}